\documentclass{appolb}
\usepackage{epsfig}
\usepackage{amssymb,longtable,amsmath,array,color,graphicx}
\def\runhead{}

\newcommand{\be}{\begin{equation}}
\newcommand{\ee}{\end{equation}}
\newcommand{\ba}{\begin{eqnarray}}
\newcommand{\ea}{\end{eqnarray}}
\newcommand{\s}{\scriptscriptstyle}
\begin{document}
% \eqsec % uncomment this line to get equations numbered by (sec.num)
\title{New Physics bounds from the combination of CKM-universality and high energy data%
\thanks{Presented at the FLAVIAnet topical workshop \lq\lq Low energy constraints on extensions of the Standard Model'' (Kazimierz, Poland), July 2009, based on \cite{Cirigliano:2009wk}.}%
}
\author{Mart\'{\i}n Gonz\'alez-Alonso
\address{Departament de F\'{\i}sica Te\`orica and IFIC, Universitat de Val\`encia-CSIC,\\
Apt. Correus 22085, E-46071 Val\`encia, Spain}
}
\maketitle
\begin{abstract}
Through an effective field theory approach, we analyze the new physics (NP) corrections to muon and beta decays and their effects on the extractions of $V_{ud}$ and $V_{us}$. Assuming nearly flavor blind NP interactions, we find that the only quantity sensitive to NP is $\Delta_{\rm CKM} \equiv |V_{ud}|^2 + |V_{us}|^2 + |V _{ub}|^2 - 1$, that receives contributions from four short distance operators. The phenomenological bound $\Delta_{\rm CKM} = (-1 \pm 6) \times 10^{-4}$ provides strong constraints on all four operators, corresponding to an effective scale $\Lambda > 11$ TeV (90\% CL). Depending on the operator, this constraint is at the same level or better than that generated by the Z pole observables.
\end{abstract}
%\PACS{13.20.-v,12.15.-y}
 
\section{Introduction}
Thanks to the precise experimental measurements~\cite{Amsler:2008zzb} and the theoretical improvements~\cite{theoreticalimprovement}, semileptonic (SL) decays of light quarks are a deep probe of the nature of weak interactions~\cite{Marciano:2007zz,Antonelli:2008jg}. In particular, the precise determination of the elements $V_{ud}$ and $V_{us}$ of the Cabibbo-Kobayashi-Maskawa (CKM)~\cite{CKM} matrix enables tests of the CKM unitarity condition\footnote{$V_{ub} \sim 10^{-3}$ contributes negligibly to this relation.} $|V_{ud}|^2 + |V_{us}|^2 + |V_{ub}|^2 =1$, at the level of $0.001$ or better. Assuming that NP contributions scale as $\alpha/\pi (M_W^2/\Lambda^2)$, this test probes energy scales $\Lambda$ on the order of the TeV, which will be directly probed at the LHC.

While the consequences of Cabibbo universality tests have been considered in some explicit Standard Model (SM) extensions~\cite{modeldependent}, our goal is to study it in a model-independent way. We have to analyze the NP contributions to the muon decay (where the $G_F$ is extracted) and to the different channels that are used to extract the product $G_F V_{ud,us}$. Currently the determinations of $V_{ud}$ and $V_{us}$ are dominated by super-allowed nuclear beta decays~\cite{Hardy:2008gy} ($V_{ud} = 0.97425(22)$) and $K_{l3}$ decays~\cite{Antonelli:2009ws} ($V_{us} =0.2252(9)$) respectively, although experimental and theoretical improvements in other channels can make them competitive in the near future.
%In light of this we set out to perform a comprehensive analysis of possible NP effects in any extraction of $V_{ud,us}$.

\section{Weak scale effective lagrangian}
\label{sect:weakscale}

In order to analyze in a model-independent framework NP contributions to both electroweak precision observables (EWPO) and beta decays we take the SM (including the Higgs) as the low-energy limit of a more fundamental theory, and more specifically we assume that: (i) there is a gap between the weak scale $v$ and the NP scale $\Lambda$ where new degrees of freedom appear; (ii) the NP at the weak scale is weakly coupled, so the electroweak (EW) gauge symmetry is linearly realized; (iii) the violation of total lepton and baryon number is suppressed by a scale much higher than $\Lambda \sim {\rm TeV}$. These assumptions lead us to an effective non-renormalizable lagrangian of the form~\cite{Buchmuller:1985jz}:
\ba
\label{eq:EFT}
{\cal L}^{(\rm{eff})} &=& {\cal L}_{\rm{SM}} + \frac{1}{\Lambda} {\cal L}_5 + \frac{1}{\Lambda^2} {\cal L}_6 + \frac{1}{\Lambda^3} {\cal L}_7 + \ldots
\ea
where ${\cal L}_n = \sum_i \alpha^{(n)}_i O_i^{(n)}$, being ${\cal O}_i^{(n)}$ local gauge-invariant operators of dimension $n$ built out of SM fields. It can be shown that under the above assumptions, there are no corrections to the SM lagrangian at dimension five, whereas seventy-seven operators appear at dimension six~\cite{Cirigliano:2009wk,Buchmuller:1985jz}, where we truncate the expansion. In order to be consistent with this truncation we will work at linear order in the NP corrections.

For the EWPO {\it and} beta decays it can be shown that we only need a twenty-five operator basis, with twenty-one $U(3)^5$ invariant and four non-invariant\footnote{We refer to the $U(3)^5$ flavor symmetry of the SM gauge lagrangian (the freedom to make $U(3)$ rotations in family space for each of the five fermionic gauge multiplets).} (we will see the usefulness of this separation later). Only nine of those operators will contribute to the beta and muon decays:
\begin{eqnarray}
 && O_{ll}^{(1)}=\frac{1}{2} (\overline{l} \gamma^\mu l) (\overline{l} \gamma_\mu l)
 ~~~~~~~~~~~~~~~~~~O_{ll}^{(3)} = \frac{1}{2} (\overline{l} \gamma^\mu \sigma^a l) (\overline{l} \gamma_\mu \sigma^a l),  \label{eq:oll} \\
 && O_{l q}^{(3)}= (\overline{l} \gamma^\mu \sigma^a l) (\overline{q} \gamma_\mu \sigma^a q), \label{eq:olq} \\
 && O_{\varphi l}^{(3)}=\! i (h^\dagger\!D^\mu \sigma^a \!\varphi)(\overline{l} \gamma_\mu \sigma^a l)+\!{\rm h.c.}, \label{eq:ohl}
 ~ O_{\varphi q}^{(3)} =\! i (\varphi^\dagger\!D^\mu \sigma^a \!\varphi)(\overline{q} \gamma_\mu \sigma^a\! q)+\!{\rm h.c.}, \label{eq:ohq} \\
&& O_{qde} = (\overline{\ell} e) (\overline{d} q)+ {\rm h.c.}, \label{eq:oqde} \\
&& O_{l q} = (\bar{l}_a e)\epsilon^{ab}(\bar{q}_b u)+ {\rm h.c.} \label{eq:olq2}
~~~~~~~~~~~ O^t_{l q} = (\bar{l}_a\sigma^{\mu\nu}e)\epsilon^{ab}(\bar{q}_b\sigma_{\mu\nu}u)+ {\rm h.c.} \\
&& O_{\varphi \varphi} = i(\varphi^T \epsilon D_\mu \varphi) (\overline{u}\gamma^\mu d)+ {\rm h.c.}~, \label{eq:ohh}
\end{eqnarray}
where only the first five are $U(3)^5$-invariant.

\section{Effective lagrangian for $\mu$ and quark $\beta$ decays}
\label{sect:gevscale}
Deriving the low-energy effective lagrangian that describes the muon and beta decays (see ref. \cite{Cirigliano:2009wk} for details) we find
\ba
%{\cal L}_{\mu \to e \bar{\nu}_e \nu_\mu}
{\cal L}_{\mu}
= \frac{-g^2}{2 m_W^2}  \! \! \! \!\! \! \! \! \! \! \! \!&& \Bigg[ \left(1 +
\tilde{v}_L
\right) \cdot \bar{e}_{\s{L}} \gamma_\mu \nu_{e\s{L}}
\ \bar{\nu}_{\mu \s{L}} \gamma^\mu \mu_{\s{L}}  \ + \
\tilde{s}_R
\cdot \bar{e}_{\s{R}} \nu_{e\s{L}} \ \bar{\nu}_{\mu \s{L}} \mu_{\s{R}} \Bigg] ~+~ h.c.~, ~
\label{eq:leffmu}\\
\label{eq:A}
\tilde{v}_L &= & 2~[\hat{\alpha}_{\varphi l}^{(3)}]_{11+22^*} - [\hat{\alpha}_{ll}^{(1)}]_{1221} - 2 [\hat{\alpha}_{ll}^{(3)}]_{1122-\frac{1}{2}(1221)}\\
\label{eq:B}
\tilde{s}_R &=& +2 [\hat{\alpha}_{le}]_{2112}~,\\
%{\cal L}_{d_j \to u_i \ell^- \bar{\nu}_\ell}
{\cal L}_{d_j}
= \frac{-g^2}{2 m_W^2} \, V_{ij}  \! \! \! \!\! \! \! \! \! \! \! \!&&\Bigg[
 \Big(1 + [v_L]_{\ell \ell ij} \Big) \ \bar{\ell}_L \gamma_\mu \nu_{\ell L} \ \bar{u}_L^i \gamma^\mu d_L^j
 \ + \ [v_R]_{\ell \ell ij} \ \bar{\ell}_L \gamma_\mu \nu_{\ell L} \ \bar{u}_R^i \gamma^\mu d_R^j
\nonumber\\
&&+ [s_L]_{\ell \ell ij} \ \bar{\ell}_R \nu_{\ell L} \ \bar{u}_R^i d_L^j
\ + \ [s_R]_{\ell \ell ij} \ \bar{\ell}_R \nu_{\ell L} \ \bar{u}_L^i d_R^j
\nonumber \\
&&+ [t_L]_{\ell \ell ij} \ \bar{\ell}_R \sigma_{\mu \nu} \nu_{\ell L} \ \bar{u}_R^i \sigma^{\mu \nu} d_L^j
\Bigg]~+~h.c.~,
\label{eq:leffq} \\
\label{eq:beta01}
V_{ij}  \cdot \left[v_{L}\right]_{\ell \ell i j} &=& 2 \, V_{ij} \, \left[\hat{\alpha}_{\varphi l}^{(3)}\right]_{\ell\ell} + 2 \, V_{im} \left[\hat{\alpha}_{\varphi q}^{(3)}\right]_{jm}^*- 2\, V_{im} \left[\hat{\alpha}_{l q}^{(3)}\right]_{\ell\ell mj} \\
V_{ij} \cdot \left[v_R\right]_{\ell \ell ij } &=& - \left[\hat{\alpha}_{\varphi \varphi}\right]_{ij} \\
V_{ij} \cdot \left[s_L\right]_{\ell \ell ij } &=& - \left[\hat{\alpha}_{l q}\right]_{\ell\ell ji}^* \\
V_{ij} \cdot \left[s_R\right]_{\ell \ell ij} &=& - V_{im}\left[\hat{\alpha}_{qde}\right]_{\ell\ell jm}^* \\
V_{ij} \cdot \left[t_L\right]_{\ell \ell ij } &=& - \left[\hat{\alpha}^t_{l q} \right]_{\ell\ell ji}^* ~.
\label{eq:beta1}
\ea

\section{Flavor structure of the effective couplings}
\label{sect:flavor}
%
%Using the general effective lagrangians (\ref{eq:leffmu}) and (\ref{eq:leffq}) for CC transitions, one can calculate the deviations from SM predictions in various SL decays and in principle a rich phenomenology is possible, because $V_{ud}$ and $V_{us}$ can be determined with high precision through different channels.
%
So far we have not made any assumption about the flavor structures in the couplings $[\hat{\alpha}_X]_{abcd}$. However flavor changing neutral current (FCNC) processes forbid generic structures if $\Lambda \sim {\rm TeV}$ and therefore we organize the discussion in terms of perturbations around the $U(3)^5$ flavor symmetry limit, where no problem arises with FCNC. 

In the $U(3)^5$-limit the expressions greatly simplify and all the NP effects can be encoded into the following redefinitions
\ba
G_F^\mu &=& (G_F)^{(0)} \, \left(1 + 4 \, \hat{\alpha}_{\varphi l}^{(3)} - 2 \,  \hat{\alpha}_{ll}^{(3)} \right)~, \\
G_F^{\rm SL}&=& (G_F)^{(0)} \, \left( 1 + 2 \left( \hat{\alpha}_{\varphi l}^{(3)} + \hat{\alpha}_{\varphi q}^{(3)}- \hat{\alpha}_{l q}^{(3)}\right)\right)~,
\ea
where $G_F^{(0)} = g^2/(4 \sqrt{2} m_W^2)$. Consequently we will have
\ba
\label{eq:vphenoFB}
V_{ij}^{(\rm pheno)}
&=& V_{ij} \left[1 +2\,\left( \hat{\alpha}_{ll}^{(3)} -\hat{\alpha}_{lq}^{(3)} -\hat{\alpha}_{\varphi l}^{(3)} +\hat{\alpha}_{\varphi q}^{(3)}\right)\right]~,
\ea
as phenomenological values of $V_{ud,us}$. This shift is independent of the channel used to extract $V_{ud,us}$ and the only way to expose NP contributions is to construct universality tests, in which the absolute normalization of $V_{ij}$ matters. Therefore the NP effects are entirely captured by the quantity
\be
\label{eq:dckm}
\Delta_{\rm CKM} \equiv |V_{ud}^{(\rm pheno)}|^2+|V_{us}^{(\rm pheno)}|^2+|V_{ub}^{(\rm pheno)}|^2 \ - \ 1 ~,
\ee
that in our framework takes the value
\be
\Delta_{\rm{CKM}}
=       4 \, \left( \hat{\alpha}_{ll}^{(3)} -\hat{\alpha}_{l q}^{(3)} - \hat{\alpha}_{\varphi l}^{(3)} + \hat{\alpha}_{\varphi q}^{(3)}  \right)~.
\label{eq:dckmnp}
\ee
%
%In specific SM extensions, the $\hat{\alpha}_i$ are functions of the underlying parameters. Therefore, through the above relation one can work out the constraints of quark-lepton universality tests on any weakly coupled SM extension.
%

The Minimal Flavor Violation (MFV) hypothesis requires that $U(3)^5$ symmetry is broken in the underlying model only by structures proportional to the SM Yukawa couplings~\cite{MFV}, and structures generating neutrino masses~\cite{Cirigliano:2005ck}. But in MFV the coefficients parameterizing deviations from the $U(3)^5$-limit are highly suppressed \cite{Cirigliano:2009wk} and so we expect the conclusions of the previous subsection to hold. The elements $V_{ij}$ receive a common dominant shift plus suppressed channel-dependent corrections.

In a more general framework the situation can be different because the channel-dependent shifts to $V_{ij}$ could be appreciable and $\Delta_{\rm CKM}$ would depend on the channels used. Work in this direction is in progress.

\section{$\Delta_{\rm CKM}$ versus precision EW measurements}
\label{sec:InvaraintAnalysis}

In the limit of approximate $U(3)^5$ invariance, we have shown that $\Delta_{\rm CKM}$ constraints a specific combination of the coefficients, that also contribute to the EWPO~\cite{Han:2004az}, together with the remaining seventeen operators that make up the $U(3)^5$ invariant sector of our TeV scale effective lagrangian. 

\begin{figure}
\includegraphics[width=12.5cm]{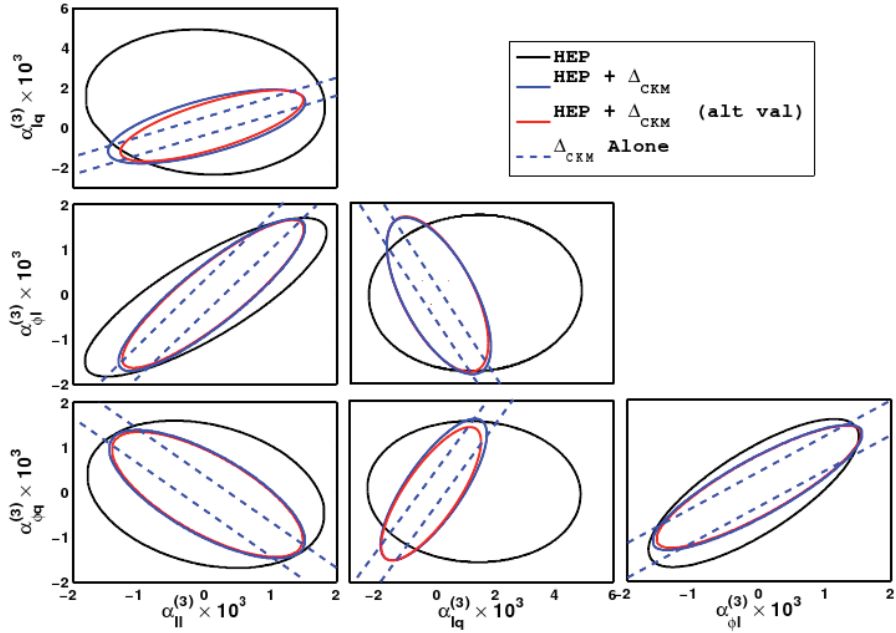}
\caption{$90\%$ C.L. regions projections, using the high energy observables (HEP), the current $\Delta_{\rm CKM}$ constraint or an alternative value of $\Delta_{\rm CKM} = -0.0025 \pm 0.0006$.%The blue dashed lines represent the section of the $90\%$ allowed regions with that planes.
\label{fig:DCKMContour} }
\end{figure}

The analysis of Han and Skiba~\cite{Han:2004az}, that studied the constraints on the same set of twenty-one $U(3)^5$ invariant operators via a global fit to the EWPO,  allows us to compare the bound on $\Delta_{\rm CKM}$ that we get from them
\be
 - 9.5 \times 10^{-3}  \ \leq \ \Delta_{\rm CKM} \ \leq \ 0.1 \times 10^{-3} \qquad (90\% \ {\rm C.L.})~,
\ee
with the direct experimental bound $|\Delta_{\rm CKM}| \leq 1. \times 10^{-3}$ ($90\%$ C.L.)~\cite{Antonelli:2009ws}. We see that EWPO leave room for a sizable non-zero $\Delta_{\rm CKM}$ and consequently we have to include the direct $\Delta_{\rm CKM}$ constraint in the global fit to improve the bounds on NP-couplings (see results in Fig.~\ref{fig:DCKMContour}). We see that the main effect is to strengthen the constraints on $O_{l q}^{(3)}$.

\begin{figure}
\hspace{0.3cm}
\includegraphics[scale=0.57]{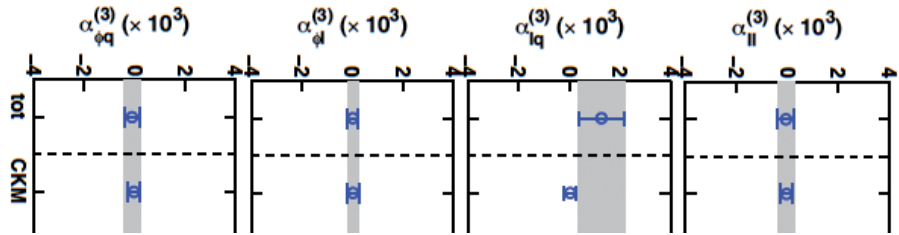}
\caption{$90 \, \%$ C.L. regions in the single operator analysis. The first row displays the constraint from EWPO and the second row those coming only from $\Delta_{\rm CKM}$.}
\label{fig:ZoomedIndOpConst}
\end{figure}

In Fig.~\ref{fig:ZoomedIndOpConst} we show the bounds if we assume a single operator dominance. For all the CKM-operators the direct $\Delta_{\rm CKM}$ measurement provides competitive constraints and in the case of $O_{lq}^{(3)}$ the improvement is remarkable.
If a non-zero $\Delta_{\rm CKM}$ is observed, in the single-operator framework it would be correlated to deviations from the SM expectation in other observables as well. These correlations have been studied in ref. \cite{Cirigliano:2009wk}.

\section{Conclusions}
\label{sec:Conclusions}

In a model-independent framework and assuming nearly flavor blind NP interactions, it has been shown that the extraction of $V_{ud,us}$ is channel independent and the only NP probe is $\Delta_{\rm CKM}$, that receives contributions from four short distance operators: $O_{ll,lq,\varphi l, \varphi q}^{(3)}$.

It has been shown that Cabibbo universality tests provide constraints on NP that currently cannot be obtained from other EW precision tests and collider measurements. The $\Delta_{\rm CKM}$ constraint bounds the effective NP scale of all four CKM-operators to be $\Lambda > 11$ TeV (90 \% C.L.), what for $O_{l q}^{(3)}$ is almost one order of magnitude stronger than EWPO-bound. Equivalently, should $V_{ud}$ and $V_{us}$ move from their current central values~\cite{Antonelli:2008jg}, EWPO data would leave room for sizable deviations from quark-lepton universality.

\section*{Acknowledgments}
%I thank the Institute for Nuclear Theory at the University of Washington for their hospitality during the completion of this work.
I thank V. Cirigliano and J. Jenkins for their help and the LANL T-2 Group for its hospitality and partial support. Work supported by the EU RTN network FLAVIAnet [MRTN-CT-2006-035482] and MICINN, Spain [FPU No. AP20050910, FPA2007-60323 and CSD2007-00042 -CPAN-].
%This work was performed under the auspices of the National Nuclear Security Administration of the U.S. Department of Energy at Los Alamos National Laboratory under Contract No. DE-AC52-06NA25396, and was supported in part by the LANL LDRD program.

%\include{biblio}

\end{document}